\documentclass{article}

\usepackage{amsmath}
\usepackage{amssymb}

\usepackage{titlesec}
\usepackage{fancyhdr}
\usepackage{booktabs}
\usepackage{graphicx}

\usepackage[labelfont=bf]{caption}
\usepackage[position=t,singlelinecheck=off]{subfig}

\usepackage[section]{placeins}
\usepackage{geometry}
\usepackage[colorlinks,
          linkcolor = black,
          anchorcolor = black,
          urlcolor = black,
          citecolor = black,
          ]{hyperref}
 
\usepackage{enumerate}

\newtheorem{myDef}{Definition}

\usepackage{float}

\providecommand{\keywords}[1]
{
  \small	
  \textbf{\textbf{Keywords:}} #1
}

\DeclareUnicodeCharacter{3002}{ }

\title{\bfseries Abnormal-aware Multi-person Evaluation System with Improved Fuzzy Weighting}
\author{Shutong Ni \\ 
\textit{School of Statistics and Applied Mathematics, Anhui University of Finance and}\\
\textit{Economics, Bengbu, Anhui, China}\\}
\date{}
\begin{document}
\maketitle







\begin{abstract}
    There exists a phenomenon that subjectivity highly lies in the daily evaluation process. Our research primarily concentrates on a multi-person evaluation system with anomaly detection to minimize the possible inaccuracy that subjective assessment brings. We choose the two-stage screening method, which consists of rough screening and score-weighted Kendall-$\tau$ Distance to winnow out abnormal data, coupled with hypothesis testing to narrow global discrepancy. Then we use Fuzzy Synthetic Evaluation Method(FSE) to determine the significance of scores given by reviewers as well as their reliability, culminating in a more impartial weight for each reviewer in the final conclusion. The results demonstrate a clear and comprehensive ranking instead of unilateral scores, and we get to have an efficiency in filtering out abnormal data as well as a reasonably objective weight determination mechanism. We can sense that through our study, people will have a chance of modifying a multi-person evaluation system to attain both equity and a relatively superior competitive atmosphere.
\end{abstract}


\maketitle

\section{Introduction}

The evaluation system has long been an indispensable part of measuring the performance of particular behaviour. For years, subjective evaluation and objective assessment have been rather separate in their respective fields. However, with the booming improvement in science and technology, these two indicators are somehow gradually intertwined and have the objective one taken the lead. Even so, subjective evaluation cannot be erased for good, on the account that it has its unique characteristics indeed, which can be generalised as minute scope, fair adaptability, low cost, and high randomness. When it comes to examinations or appraisals, expertise revision of contributions, personnel recruitment, project bidding, judgments on equipment's function or merchandise's quality, and even government's policymaking, subjective assessment operates in every tiny aspect of the society, paving the way for its unremitting upswing.

Practically, the two evaluation strategies have their leanings. To minimise the repercussion of their defects, under a particular circumstance, there are scholars wedded to incorporating the two assessment methods together, in the hope that the results can be much fairer as well as more reliable \cite{sleep,TV_Interfaces,vehicle}. Nevertheless, in many scenarios, objective evaluation data is hard to obtain, and many a strength of subjective evaluation make it a more practical means. A simple way to increase the credibility of the evaluation is to summarise the information of multiple reviewers, which will lead to the inconsistency of results so that a significant number of researchers address themselves into how to integrate various information to make the ultimate review authentic as much as possible. To avert a mixture of standards, experts should conduct an evaluation with respect to various indices in different situations. Xing \cite{RufeiXing} proposes correlation analysis \cite{correlation} to measure the reliability of reviewers, screening out the discrepant values, and, at the same time, streamlining the assessment indices that possess a strong correlation with each other. Regarding that different reviewer has different standards over objects, Xing \cite{RufeiXing} choose to apply Fuzzy Analysis Hierarchy Process(FAHP) \cite{FAHP} to the problem, in the hope of a comparable and definite weight factor for each evaluation index. That is to say, by forming a fuzzy judgment matrix that consists of experts' appraisals, we can determine the weight factors in the assessment system.

Nevertheless, the pose of a specific assessment index will also be affected both subjectively and professionally, and on a certain condition, we cannot even put forth a scientific index, for instance, when teachers in schools rate their students, the only factor worth referring to is the final score. Thus, our research will prioritize the questions that concern information-limited problems, which subsequently introduce us to refine the disadvantages of subjective evaluation as below:

To start with, the evaluation standard differs from person to person, so there is no absolute right or wrong, and every authority has his/her own precept and preference. In other words, for a certain wide range of scoring standards, diverse experts have enormous differences in the understanding of the evaluation standards and the grasp of the assessment scales, which simultaneously give rise to a conspicuous contrast. The second lies in the psychological impact of each expert during the evaluation process. When scoring, the experts will inevitably be influenced more or less by the scores of other students he has given. That is to say, subsequent scores will be subject to all of the previous scoring results, which conduce to the essential variation.

According to the imperfections that exist in the subjective evaluation, our study intends to modify the multi-person subjective evaluation method, and we endeavour to provide resolutions in the design of assessment procedure as well as assessment approaches, making evaluation results and objective facts converge as much as possible. In this case, we may help stamp out the bias and constraints of individual evaluation and demonstrate impartiality as well as authority, so as to shape a superior competitive atmosphere.

In our research process, we use two-step screening as a quick start to examine the anomalous data. Given that the standard Q-test method \cite{Qtest} and 3-$\sigma$ principle \cite{threesigma} fail to work well when the samples are inadequate, we can utilize these methods for a rough selection, and then apply Kendall-$\tau$ Distance to examine the data winnowed out for the sake of advancing the secondary screening procedure. In light of the drawbacks that original Kendall-$\tau$ Distance can hardly fully contemplate the differences among values of scores, we rework the idea as score-weighted Kendall-$\tau$ Distance and regard it as an objective function, screen out the abnormal data which will result in a minor decrease in the objective. Moreover, still in the data preprocessing stage, we propose to mitigate the discrepancy of scores given by the same reviewer between two classes through hypothesis testing. Later, considering the fuzzy relationships among reviewers' judging criteria, we interpret the outcomes from different experts as different judging indices of objects evaluated, so as to use the fuzzy synthetic evaluation model\cite{FSEwater1, FSEwater2, FSErisk1, FSErisk2}. When weighing those experts' reviews, we not only take the weight stemmed from the fuzzy synthetic evaluation into consideration, but also calculate another type of weight derived from the scale of anomalous data excluded, which can tell the reliability of the evaluation. Coupling the two weights with each other, we take the average as the final weight for the index, thereby measuring the accuracy of evaluation more efficiently.

In the following sections, we first present the formulation of the entire problem. And then, we narrate the formation process of our evaluation method, including the two-stage screening method, hypothesis test, and fuzzy synthetic evaluation method. Subsequently, we conduct experiments and show the effectiveness of our methods. Furthermore, we ultimately draw a conclusion of the questions and elucidate the notion of our research and modification for a more equitable evaluation system.

\section{Problem Formulation}
\label{sec2}
In this section, we introduce standard problem formulation that we aim to tackle via our proposed model. Moreover, we also specify certain assumptions which contribute to the completeness of our definition.

Let $g_{ij}^k$ denotes the grade given by reviewer $i$ to student $j$ in class $k$, where $i=1,2,...,R, \ j=1,2,...,S_k, \ k = 1,2,...,C.$ Let $\pi_i$ denotes the ranking of all students, whose papers are reviewed by reviewer $i$, in descending order of their grades.

We mainly focus on two representative assessment situations. The first is that we are given one class with $S$ students whose papers need to be graded by $R$ reviewers.
The second situation is that we are given $k$ classes, where $k\ge 2$, and each of the $R$ reviewers must grade all the classes. In both cases, the task is to estimate the actual grades of all the students and determine their final rankings as fair and objective as possible, without knowing about the evaluation principles or preferences of each reviewer.

It is worth mentioning the following assumptions that help avoid certain intricate controversies:
\begin{enumerate}[(1)]
 \item Each reviewer grades papers under the same external conditions.
 \item All papers are kept secret before reviewing.
 \item Reviewers are not allowed to discuss with each other.
 \item Students' rankings are only determined by their final synthesized scores. 
\end{enumerate}

\section{The Proposed Evaluation System}

 \subsection{Anomaly Detection}
 Since the only difference between the two situations is the number of classes, we decide to tackle this at the end of the analysis. For the screening of the abnormal data, the most common method is indubitably the Trimmed Mean, which strikes out a certain proportion of the highest and lowest scores, and averages the remaining ones. However, in our research, we actually have only a small amount of reviewers for the evaluation task, where the Trimmed Method cannot operate to its full potential and has low robustness \cite{trimmed}. On the one hand, under the premise that students have only a few scores, removing high scores or low scores can reduce extreme data to a certain extent, but it also causes a significant loss of data. In this case, for some superficially extreme scores, we decide to winnow them out instead of directly discarding them. This effectively considers that despite the extremity of the lateral comparison among scores given by different reviewers, this kind of extremity can be rather valuable in the whole rankings given by a specific reviewer. Taking this factor into consideration and then designing a rational objective function, we can reach out to a relatively optimal two-step screening method. On the other hand, each reviewer may generally rate students from different classes high or low due to various evaluation standards, and there remains a certain contrast among scoring intervals. Therefore, the scores given by reviewers do not comply with the normal distribution of their true average score so the horizontal comparison can be meaningless. As a consequence of that, If we substitute students' scores with their rankings, this new indicator can also make the two-step screening method well-performed.

 \subsubsection{Rough Screening}
 \label{anomalyI}
 There are plenty of typical means to filter outliers, such as 3-$\sigma$ principle, quantile method and Q-test, etc. However, in terms of our research, we notice that there are not enough samples, because each student have gained scores only from few reviewers, and at the same time, the degree of anomaly fails to reach the standards of the methods mentioned above. For the second situation, after we transform their original scores into rankings as discusses above, it will face the same trouble.

 Aware of this problem, we alter our perspective from designing a efficient one-round screening method to a multi-stage method. In the first step, we roughly calculate the average and variance of students' scores, then subsume data into set $\mathcal{G}$ if its deviation from the average is greater than $\alpha$ times of its variance, where $\mathcal{G}$ is called anomaly set and $\alpha$ is a hyperparameter required to be fine-tuned. The second step is much more pivotal and will be specified in the following section.

 \subsubsection{Score-weighted Kendall-\texorpdfstring{$\tau$}{Lg} Distance}
 \label{anomalyII}
 To conduct the second step of winnowing out abnormal data, we would like to introduce the definition of Kendall-$\tau$ Distance \cite{kendall} to you for a quick start. 
 \begin{myDef}
     We define the Kendall-$\tau$ Distance between $\tau_{1}$ and $\tau_{2}$ as below:
     \begin{align}
         \delta_K (\tau_1,\tau_2)= \sum_{u \succ_{\tau_1} v} \mathbb{I}[\{v \succ_{\tau_2} u\}]
     \end{align}
     \label{def}
 where $\{\cdot\}$ denotes certain incidents, $\mathbb{I}[\cdot]$ denotes the indicative function, and $x \succ_{\tau} y$ denotes that student $x$ is ranked higher than student $y$ in rank $\tau$.
 \end{myDef}

 That is to say, we can draw the sequence of students directly through their scores. According to definition \ref{def}, we can sense that if two reviewers' reviews on a particular student differ a great deal, then the Kendall-$\tau$ Distance will be relevantly more considerable.

 Having a deeper insight into the problem, the given definition of Kendall-$\tau$ Distance practically remains some drawbacks because it simply contains the sequencing of two students but overlooks the specific difference of scores. However, in our research, the problem we have encountered has siccar grades statistics, so that we can make a modification and obtain a score-weighted Kendall-$\tau$ Distance:

 \begin{align}
     \delta_{SK} (\pi_k,\pi_l)= \sum_{u \succ_{\pi_k} v} 
     \frac{[(g_{ku} - g_{kv}) + (g_{lv} - g_{lu})]}{2} \cdot
     \mathbb{I}[\{v \succ_{\pi_l} u\}]\cdot \mathbb{I}[\{u,v \in \pi_k\}\cap \{u,v \in \pi_l\}]
     \label{scoredKendall}
 \end{align}
 where $\pi_{k}$ and $\pi_{l}$ are rankings given by reviewer $k$ and $l$ respectively, and $ u \in \pi_{k}$ denotes that the score of student $u$ from reviewer $k$ has not been filtered out. We hope that the following objective function will become as small as possible after we winnow out the target data.
 
 \begin{align}
     \mathcal{L}(\pi_1,\pi_2,\pi_3) = \delta_{SK} (\pi_1,\pi_2) + \delta_{SK} (\pi_2,\pi_3) + \delta_{SK} (\pi_3,\pi_1)
     \label{lossfunc}
 \end{align}

 Let us respectively consider how much the objective function $\mathcal{L}$ will decline by removing a subset of the abnormal score set $\mathcal{G}$. It will become an exponential time complexity algorithm, which is pretty inadvisable. Hence, we apply a greedy methodology that only needs to figure out how much the objective function will decline when any of the abnormal scores $g_{ij} \in \mathcal{G}$ is deleted. Here, we do not necessarily use the formula(\ref{lossfunc}) for calculation every single time, but merely compute the value below: 

 \begin{align}
     \Delta_{ij} = \sum_{k \not= i} \sum_{v} 
     \frac{[|g_{ij} - g_{iv}| + |g_{kv} - g_{kj}|]}{2} \cdot
     \mathbb{I}[\{j \succ_{\pi_i} v,\ v \succ_{\pi_k} j\} \cup 
     \{j \prec_{\pi_i} v,\ v \prec_{\pi_k} j\}]. 
     \label{decrease}
 \end{align}

 After sorting $\{ \Delta_{ij} \}$ in descending order, delete the largest $d$ data in turn where the size of $d$ depends. Note that when some anomalies have been blanked out, theoretically, the corresponding $\{ \Delta_{ij}^{'} \}$ should be recalculated for the abnormal data left, and the process can no longer be implemented using the formula(\ref{decrease}) but should be added with a factor which is similar to the indicative function in formula (\ref{scoredKendall}). Yet, the actual sequence of $\{ \Delta_{ij}^{'} \}$ is exactly the same compared with $\{ \Delta_{ij} \}$. Therefore, getting rid of the anomalies greedily is reasonable to a degree. The formula(\ref{decrease}) seems to be a bit complicated, but it can be computed at a rapid pace in Python language.

 \subsection{Fuzzy Synthetic Evaluation}
  
  Based on the problem analysis in Part 2, after the screening of abnormal data, we need to synthesize all the information to determine the final score of each student. Inspired by the methods in \cite{FSEsame1, FSEsame2}, we cleverly adapt the fuzzy synthetic evaluation method to our less informatic problem setting, treating each reviewer as an evaluation index. We have the observation matrix $G = (g_{ij})_{1\le i \le m, 1 \le j \le n},$ where m is the number of reviewers and n is the number of students. Since the scores given by each reviewer can be understood as a benefit indicator, we can establish a fuzzy benefit matrix $B = (b_{ij})_{1\le i \le m, 1 \le j \le n},$ where
 \begin{align}
     b_{ij} = \frac{g_{ij} - \min_{j}g_{ij}}{\max_{j}g_{ij}-\min_{j}g_{ij}}.
 \end{align}
 It should be noted that, in accordance with Section 4.1, some $g_{ij}$ has been screened out. We do not fill in the blanks for those missing values but choose to ignore them, which means that if $g_{ij}$ is missing, $b_{ij}$ is also missing in matrix $B$.

 Coming up then, we need to establish the weight $w_i, \ 1 \le i \le m$ of each reviewer. Firstly, the coefficient of variation is adopted to fully consider the influence of the size of evaluation intervals on the degree of differentiation of students. The coefficient of variation corresponding to each reviewer is calculated according to the following formula, aiming to fully consider the unknown influence of the size of evaluation intervals on ranking:
 \begin{align}
     v_i = \frac{\sigma_i}{\overline{g}_i}, \ i = 1,...,m
 \end{align}
 where
 \begin{align}
     \overline{g}_i &= \frac{1}{n} \sum_{i=1}^{n} g_{ij}, \\
     \sigma_i &= \sqrt{ \frac{1}{n-1} \sum_{i=1}^{n} (g_{ij} - \overline{g}_i)^2 }, \ j = 1,2,...,n.
 \end{align}
 Note that when calculating $\sigma_i$ and $\overline{g}_i$, the sample size is taken as the number of scores given by the evaluation reviewer i, excluding the screening ones. Then, $v_i$ is normalized to obtain the first component of the weight:
 \begin{align}
     w_i^{(1)} = \frac{v_i}{\sum_{i=1}^m v_i}, \ i = 1,...,m
 \end{align}

 Secondly, we consider the credibility of each reviewer and use $n_i$, the number of reviewer $i$'s screened scores, to obtain another part of the weight:
 \begin{align}
     w_i^{(2)} = \frac{\sum_{i=1}^m n_i}{n_i}, \ i = 1,...,m
 \end{align}

 Finally, the weight of each reviewer is determined by the following formula:
 \begin{align}
     w_i = \frac{w_i^{(1)}+w_i^{(2)}}{2}, \ i = 1,...,m
 \end{align}
 Then $F_j = \sum_{i=1}^m w_i * b_{ij}$ is calculated for student $j$, and their ranking can be obtained after sorting $\{F_j\}$.

 Actually, the ultimate goal of our calculations is to determine the ranking of students. Now that the ranking has been obtained, If we want to output the final score, we only need to select a reference value, for example, $ \min_j g_{1j}$, and add $F_j (\max_j g_{1j}-\min_j g_{1j})$ on this basis. Then the final score $f_j$ of student $j$ is:
 \begin{align}
     \label{finalscore}
     f_{j} = F_j (\max_j g_{1j}-\min_j g_{1j}) + \min_j g_{1j}.
 \end{align}

 \subsection{Normal Hypothesis Tests}
 \label{hypotest}
 For the second situation, we propose to perform the normal hypothesis test\cite{statinfer} before screening abnormal data. Specifically, we test the mean and variance of the grades given by one reviewer to two classes. We assume that the grades given by marker $i$ to class $k$ follow the normal distribution $N(\mu_i^k, \sigma_i^k), \ i=1,...,5, \ k=1,2,$ we need to test the following two questions:
 \begin{enumerate}[(1)]
     \item $H_0: \ \mu_i^2 - \mu_i^1 = 0 \leftrightarrow H_1: \ \mu_i^2 - \mu_i^1 \not= 0$
     \item $H_0: \ \frac{\sigma_i^2}{\sigma_i^1} = 1 \leftrightarrow H_1: \ \frac{\sigma_i^2}{\sigma_i^1} \not= 1$
 \end{enumerate}
 Set the confidence level to $\alpha$. Note that $\mu_i^k$ and $\sigma_i^k$ are unknown, so that rejection region of the test for question (1) and question (2) are respectively as below:
 \begin{enumerate}[(1)]
     \item 
     \begin{align*} 
         D_1 = \{(g_{i1}^1,...,g_{in_1}^1;g_{i1}^2,...,g_{in_2}^2): \bigg| \frac{G_i^2 - G_i^1}{S_i^w} \sqrt{\frac{n_1 n_2}{n_1+n_2}} \bigg| > t_{n_1+n_2 - 2}(\frac{\alpha}{2})\}, 
     \end{align*}
     where
     \begin{align*}
         G_i^k &= \frac{1}{n_k}\sum_{j=1}^{n_k}g_{ij}^k, \\
         (S_i^w)^2 &= \frac{1}{n_1+n_2-2} [(n_1 - 1)(S_1)^2 + (n_2 - 1)(S_2)^2], \\
         (S_i^k)^2 &= \frac{1}{n_k-1} \sum_{j=1}^{n_k} (g_{ij}^k - G_i^k)^2, \ i=1,...,5, \ k=1,2.
     \end{align*}

     \item 
     \begin{align*} 
         D_2 = \{(g_{i1}^1,...,g_{in_1}^1;g_{i1}^2,...,g_{in_2}^2): \frac{(S_i^2)^2}{(S_i^1)^2} < F_{n_2-1,n_1-1}(1-\frac{\alpha}{2})
         \text{ or } \frac{(S_i^2)^2}{(S_i^1)^2} > F_{n_2-1,n_1-1}(\frac{\alpha}{2})\}.
     \end{align*}
 \end{enumerate}
 All the statistics mentioned above can be obtained from existing data. Based on the results of the two tests as well as the practical meaning of confidence rate, we scale the scores differently:

 \begin{enumerate}
     \item If both of the test results are rejection, we apply the following transformation to the scores of class 2 given by the reviewer $i$:$$g_{ij}^2 = (1-\alpha)\left[\frac{S_i^1}{S_i^2}(g_{ij}^2 - G_i^2) + G_i^1\right] + \alpha g_{ij}^2 .$$
     \item If the result of question 1 is rejection and that of question 2 is acceptance, we apply the following transformation to the scores of class 2 given by the reviewer $i$: $$g_{ij}^2 = (1-\alpha)(g_{ij}^2 - G_i^2 + G_i^1) + \alpha g_{ij}^2 .$$
     \item If the result of question 2 is rejection and that of question 1 is acceptance, we apply the following transformation to the scores of class 2 given by the reviewer $i$: $$g_{ij}^2 = (1-\alpha) \left[ \frac{S_i^1}{S_i^2} \cdot (g_{ij}^2-G_i^2) + G_i^2 \right] + \alpha g_{ij}^2.$$
     \item If both of the test results are acceptance, we keep them unchanged.
 \end{enumerate}

 After rescaling, We can merge the two classes' grades into one table, then conduct abnormal data screening and fuzzy synthetic evaluation.

 \section{Experiments}
 In this section, we conduct two experiments that testify to the effectiveness of our methods. We will specifically describe the dataset we use and present the results in detail. The detailed implementation codes are available at: \url{https://github.com/Ciao-Yvette/Multi-person-Evaluation-System}

 \subsection{Settings}
 For each aforementioned situation in section \ref{sec2}, we collect two corresponding datasets for evaluating the effect of our proposed method.
 Table \ref{data} summarizes some critical statistics of the datasets.
 
 \begin{table}[h]
 \caption{Statistics for the two datasets.}
 \label{data}
 \centering
 \begin{tabular}{ccc}
     \toprule
     Statistics & Dataset I & Dataset II\\
     \midrule
     reviewer number($R$) & 3 & 5 \\
     class number($C$) & 1 & 2 \\
     student number($S$) & 107 & 61,50 \\ 
     \bottomrule
     \end{tabular}
 \end{table}
 Notably, all the scores are rated following the percentage system, ranging from 0 to 100, preventing the uncertainties introduced from other level ranking systems. Apart from that, we do not have knowledge of the reviewers' criteria for judging.

\subsection{Results}
 \subsubsection{Experiment I}

 In the first scenario, the intervals of the grades given by the three reviewers are [65, 95], [62, 99], [60, 98], which are almost identical. Therefore, there is no need to use the student's ranking instead of grades to filter out abnormal data.

 We first apply the method proposed in section \ref{anomalyI} to screen out 33 anomalous scores. After that, we calculate the quantity for each anomalous score based on equation (\ref{decrease}) and sort the results in decreasing order as below:
 {(91, 3): 2667.0, (57, 3): 1811.5, (6, 3): 1503.5, (53, 3): 1447.5, (7, 3): 1444.5, (62, 3): 1429.5, (60, 3): 1384.5, (94, 1): 1361.0, (8, 2): 1121.5, (68, 3): 1120.0, (74, 3): 1104.0, (82, 2): 1017.0, (90, 2): 952.0, (102, 1): 871.0, (71, 3): 831.0, (48, 3): 816.0, (69, 3): 764.5, (27, 2): 757.5, (20, 3): 731.0, (19, 1): 726.0, (25, 3): 715.0, (47, 3): 696.0, (86, 1): 658.5, (33, 3): 648.0, (73, 3): 645.5, (89, 2): 642.5, (58, 3): 625.0, (51, 3): 624.0, (84, 3): 571.5, (17, 2): 548.5, (2, 1): 452.0, (78, 1): 334.0, (97, 2): 128.0 | ``$(j,i): x$" denotes $\Delta_{ij} = x$}。
 
 We set the confidence to 0.80, that is, select the first $33 \times 0.80\% \approx 26$ elements in collection $\{ \Delta_{ij} \}$ and take their corresponding $g_{ij}$ as the final anomalous scores: $\mathcal{G}=$ \{(91, 3): 98, (57, 3): 65, (6, 3): 60, (53, 3): 68, (7, 3): 67, (62, 3): 67, (60, 3): 63, (94, 1): 95, (8, 2): 85, (68, 3): 70, (74, 3): 63, (82, 2): 75, (90, 2): 85, (102, 1): 85, (71, 3): 74, (48, 3): 70, (69, 3): 71, (27, 2): 69, (20, 3): 72, (19, 1): 88, (25, 3): 62, (47, 3): 74, (86, 1): 90, (33, 3): 61, (73, 3): 73, (89, 2): 95 | ``$(j,i): x$" denotes $g_{ij} = x$\}。

 The next step is to proceed with the fuzzy synthetic evaluation. The observation matrix can be directly obtained from the dataset, and simple calculation leads to the fuzzy benefit matrix. When these are all done, we can calculate the two types of weights:
 \begin{align*}
     w^{(1)} = (0.325675, 0.377674, 0.296651) \\
     w^{(2)} = (0.349153, 0.345763, 0.305085) \\
     w = (0.337414, 0.361718, 0.300868)
 \end{align*}

 Finally, we calculate the final score using the formula (\ref{finalscore}), and the quicksort algorithm \cite{quicksort} can be harnessed to accelerate ranking. Because the weights are all decimals, we assume the final score should be kept to two decimal places. Students with the same score are regarded as the same ranking. For the specific final score table, please refer to Supplementary 1.

 \subsubsection{Experiment II}
 There are apparent differences between the evaluation intervals of the five teachers in the second dataset. As mentioned above, we propose to use the students' initial ranking as the score to screen out outliers and then conduct fuzzy synthetic evaluation and other operations. However, the greatest problem here is to mitigate the deviation of evaluation intervals. Thus we are supposed to first implement the method in section \ref{hypotest}. After rescaling, we can fairly compare the grades of different classes; thus, fuzzy synthetic evaluation can be conducted to get final scores and rankings. For the sake of emphasis, we do not present the results of the screening operation. The final scores are presented in Supplementary 2.

 \section{Discussions}
 \begin{enumerate}[(1)]
     
     \item When performing the screening of abnormal data, slightly different approaches are used in the two scenarios respectively, but these methods cannot wholly override all possible abnormalities. Suppose multiple methods are used to eliminate the outliers in data for a specific problem, and the data appears to be an exception in all methods. It can eventually determine the data is an abnormal one.
     \item The Fuzzy Synthetic Evaluation(FSE) method is a comprehensive evaluation method based on the problems of fuzziness and uncertainty in the evaluation criteria, evaluation factors, and the problem of difficulty in quantifying qualitative indicators. It can express a fuzzy object with a precise number so that the evaluation of fuzzy events is scientific and reasonable. Nevertheless, the application of the FSE method usually brings about relatively high subjectivity in the determination of indicators, fuzzy relation matrix, weight, etc. Besides, there is no clear or systematic method for determining the membership function, which conspires to a specific difference in results. 
 \end{enumerate}

\section{Conclusions} 

The evaluation problem we concentrate on in our study is based on the fact that for schooling and appraisals, there exists a certain range of situations where no uniform standard is contained. In those cases, assessment results appear to be entirely subjective and divergent. It turns out that the evaluation problem has an inextricable connection with people's daily lives, and this is why our research is intended to center on this question and make an expansion. Our research proposes a modified score-weighted Kendall-$\tau$ Distance as the judging criterion, adopts FSE as well as Normal Hypothesis Test to be the principle investigating methods, and uses Python and Matlab as auxiliary tools. Under the auspices of fundamental scientific materials, we ultimately get to winnow out anomalies and then synthesize different factors for a comprehensive evaluation, culminating in a relatively equitable judging system.



\section*{Conflict of interest}

The authors declare there is no conflicts of interest.

\bibliographystyle{plain}
\bibliography{mybibfile}

\newpage

\section*{Supplementary}
\section*{The Final Scores and Rankings}
\subsection*{1. Scenario I}
\label{supp1}
\begin{figure}[ht]
 \centering
 \includegraphics[width=130mm]{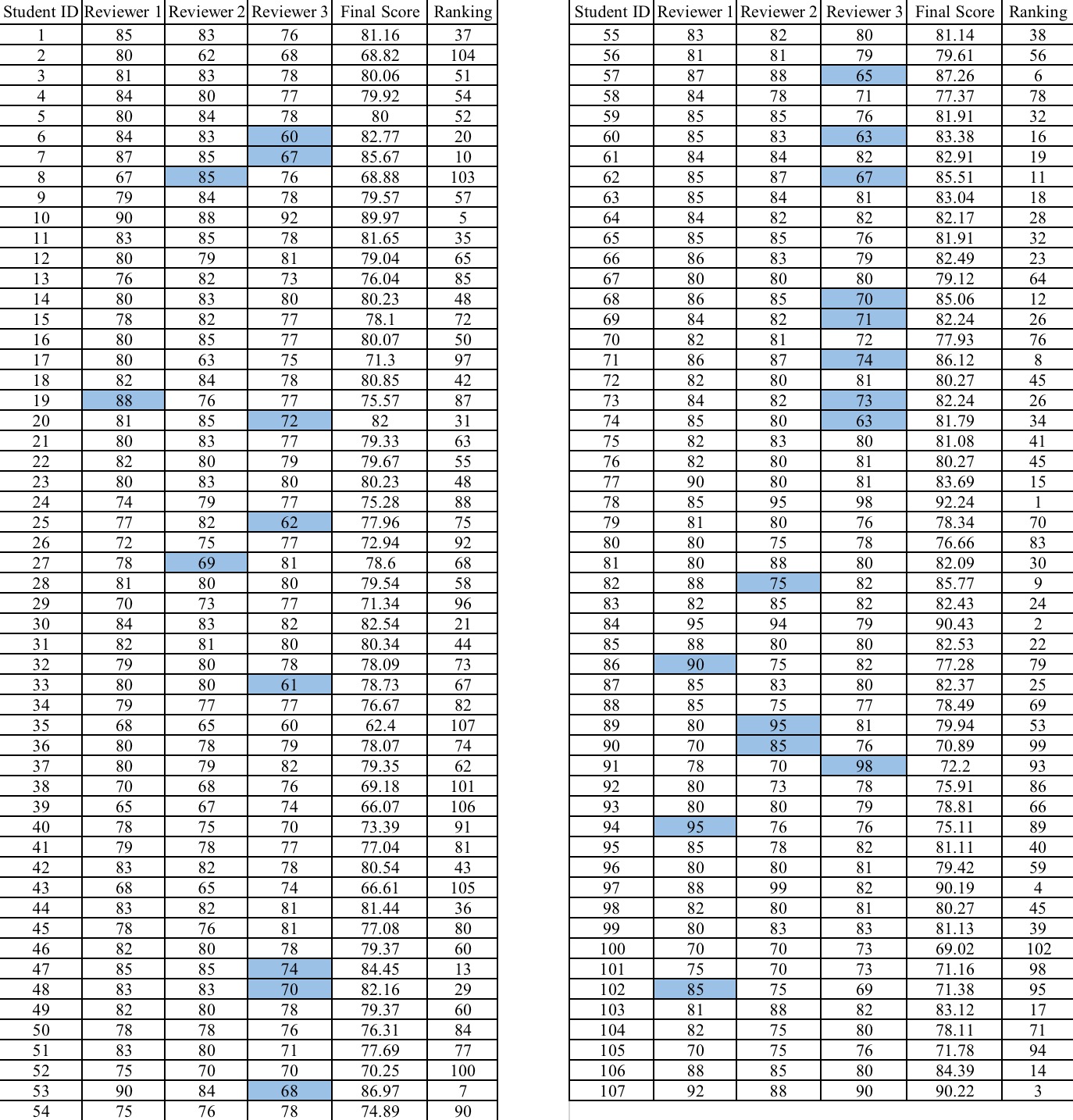}
 \caption{The final scores and rankings of all the students in both classes in scenario I.}
 \label{abnormal}
\end{figure}

The data marked blue in Figure \ref{abnormal} are abnormal.

\subsection*{2. Scenario II}
\label{supp2}
\begin{figure}[ht]
     \centering
     \includegraphics[width=75mm]{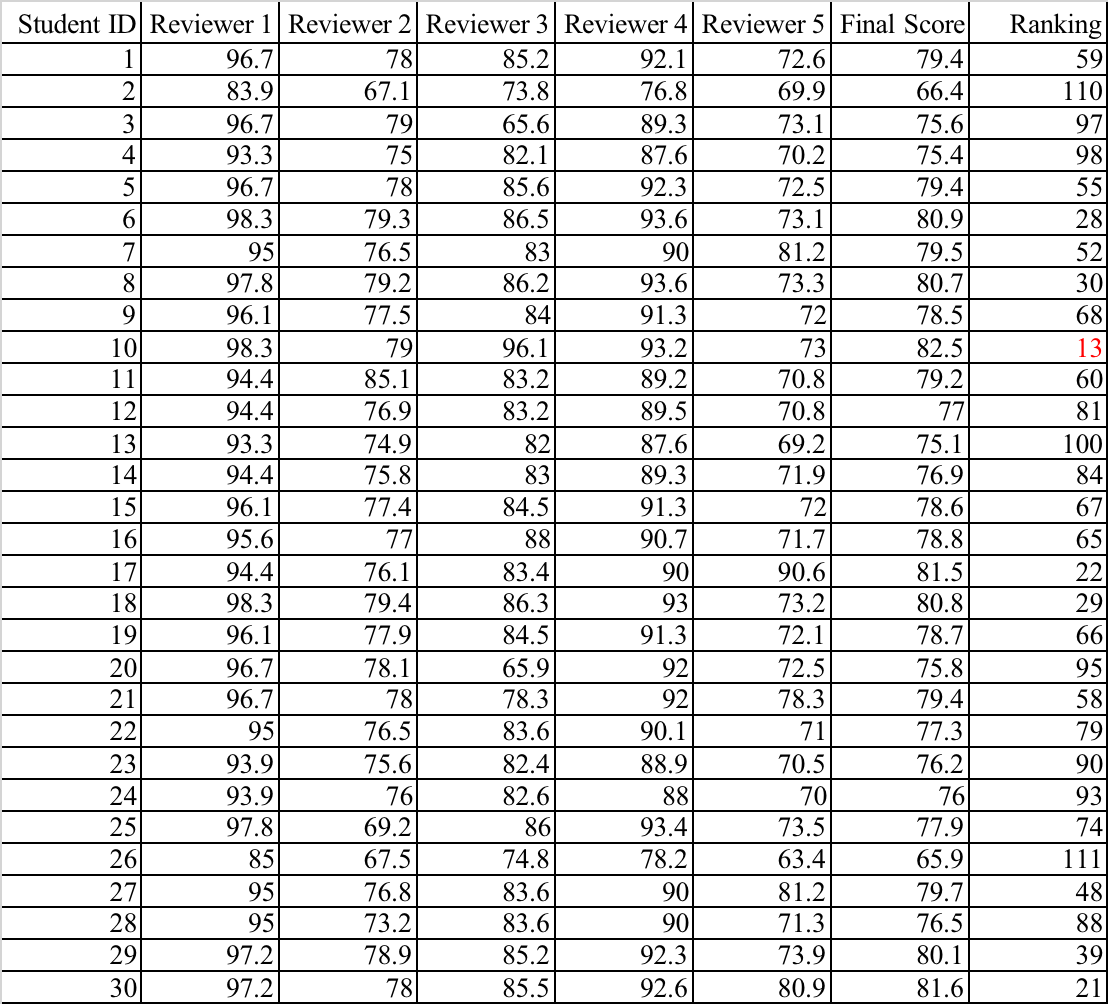}
     \includegraphics[width=75mm]{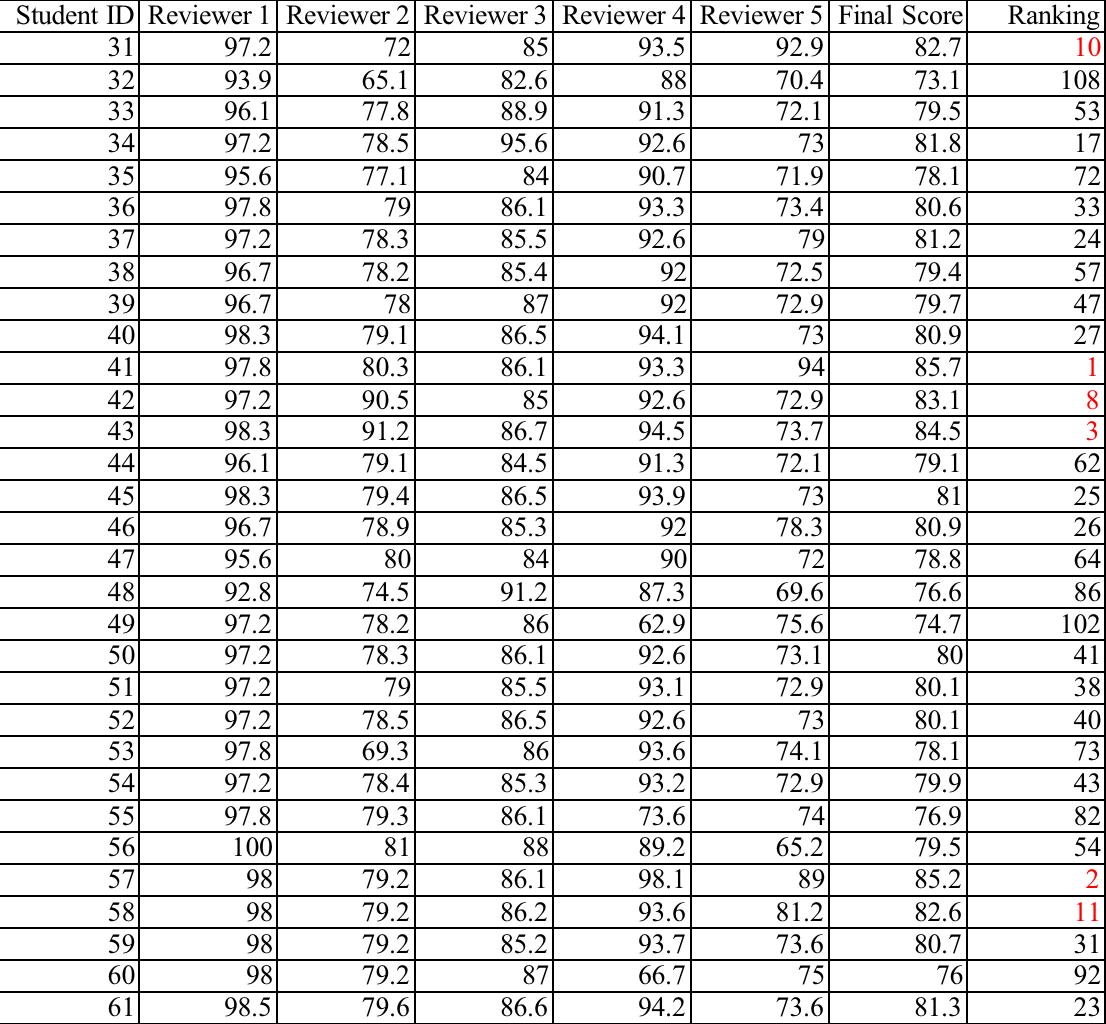}
     \caption{The final scores and rankings of the students in class 1. Rankings are calculated among all the students in both classes.}
 \label{class1}
\end{figure}

\begin{figure}[ht]
 \centering
 \includegraphics[width=75mm]{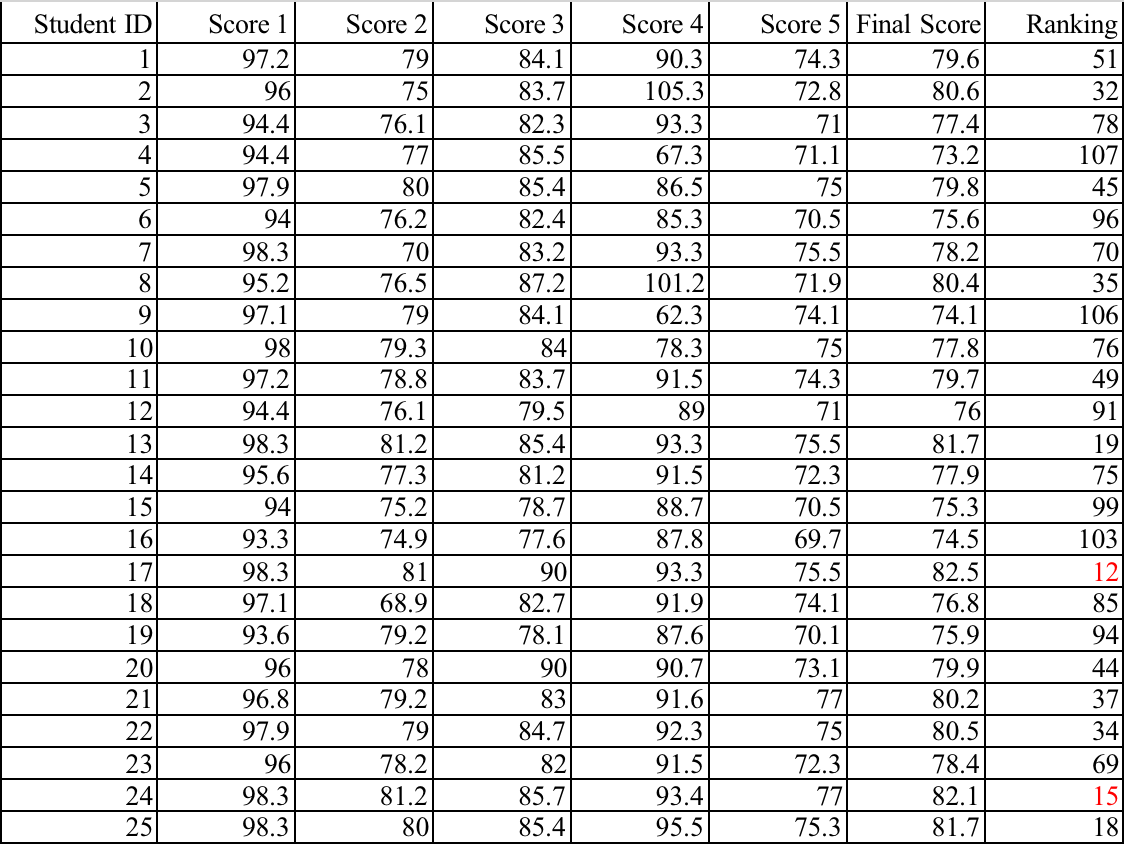}
 \includegraphics[width=75mm]{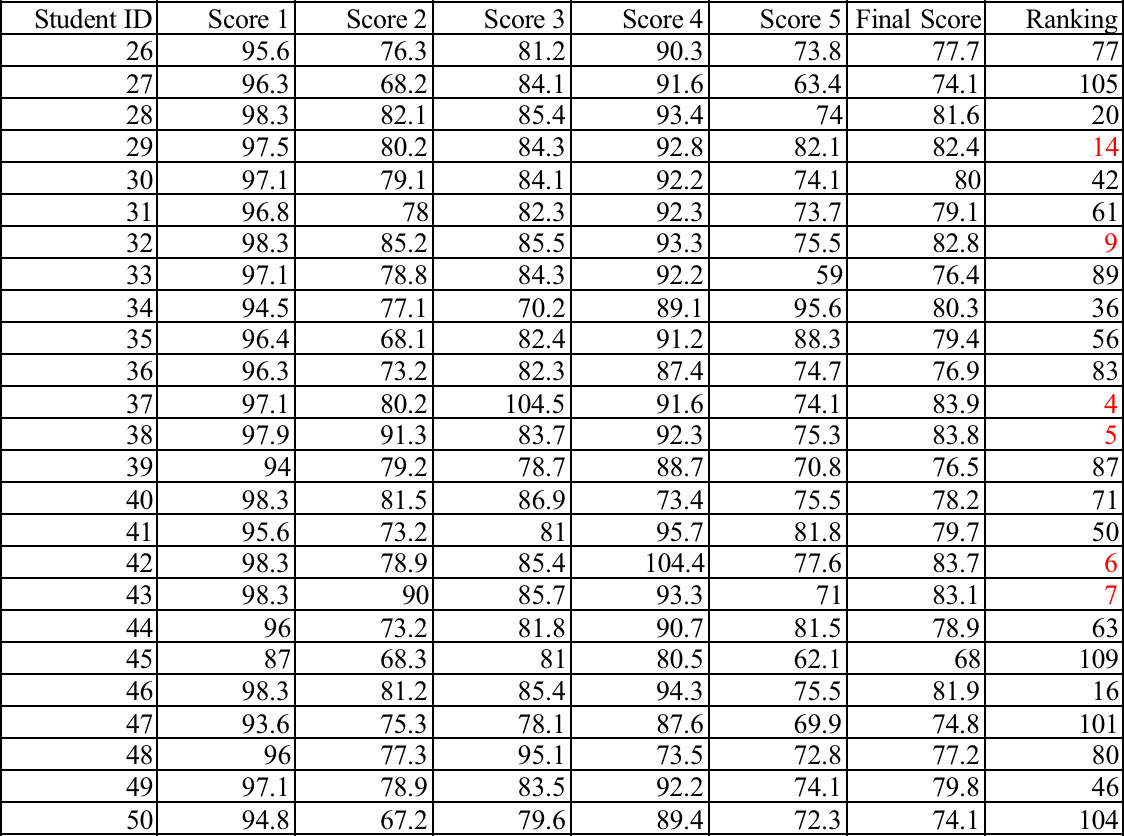}
 \caption{The final scores and rankings of the students in class 2. Rankings are calculated among all the students in both classes.}
 \label{class2}
\end{figure}

The data marked red in Figure \ref{class1} and Figure \ref{class2} are the rankings of the top 15 students.

\end{document}